\input{aipcheck}

\documentclass[
    ,final            
  ]
  {aipproc}

\layoutstyle{6x9}

\begin{document}

\title{Dark Matter Searches with GLAST}


\author{Lawrence Wai \\}{
  address={Stanford Linear Accelerator Center \\ 2575 Sand Hill Road
\\ Menlo Park, CA 94025 \\ U.S.A.  \\ {~} \\
{\em representing the GLAST LAT Collaboration}}  }

\begin{abstract}
Indirect detection of particle dark matter relies upon pair
annihilation of Weakly Interaction Massive Particles (WIMPs), which is
complementary to the well known techniques of direct detection
(WIMP-nucleus scattering) and collider production (WIMP pair
production).  Pair annihilation of WIMPs results in the production of
gamma-rays, neutrinos, and anti-matter.  Of the various experiments
sensitive to indirect detection of dark matter, the Gamma-ray Large
Area Space Telescope (GLAST) may play the most crucial role in the
next few years.  After launch in late 2007, The GLAST Large Area
Telescope (LAT) will survey the gamma-ray sky in the energy range of
20MeV-300GeV.  By eliminating charged particle background above 100
MeV, GLAST may be sensitive to as yet to be observed Milky Way dark
matter subhalos, as well as WIMP pair annihilation spectral lines from
the Milky Way halo.  Discovery of gamma-ray signals from dark matter
in the Milky Way would not only demonstrate the particle nature of
dark matter; it would also open a new observational window on galactic
dark matter substructure.  Location of new dark matter sources by
GLAST would dramatically alter the experimental landscape; ground
based gamma ray telescopes could follow up on the new GLAST sources
with precision measurements of the WIMP pair annihilation spectrum.
\end{abstract}

\maketitle


\section{Introduction}

Some of the most exciting discoveries in the history of physics may be
just around the corner.  In the next 5-10 years the particle nature of
dark matter may very well be discovered by colliders, direct
detection, and/or indirect detection, if the dark matter is composed
primarily of Weakly Interacting Massive Particles (WIMPs).  Of course,
there are other possibilities for particle dark matter, as summarized
in the following table:

\begin{table}[h]
\begin{tabular}{lcrc}
\hline
  & \tablehead{1}{c}{b}{production mechanism}
  & \tablehead{1}{c}{b}{mass scale}
  & \tablehead{1}{c}{b}{observation methods}   \\
\hline
axions & non-thermal & $\approx10^{-5}$~eV & radio wave conversion \\
neutrinos & big-bang thermal & $\approx0.1$~eV & large scale structure \\
WIMPs & big-bang thermal & $\approx100$~GeV & direct/indirect detection,\\
&&& collider production\\
\hline
\end{tabular}
\caption{non-baryonic particle dark matter candidates}
\label{tab:a}
\end{table}

The WIMP hypothesis is the favorite, mainly because of the simple
big-bang thermal production mechanism, but also because WIMPs are
accessible to colliders, direct detection, and/or indirect detection.
These complementary experimental methods will provide crucial clues in
the exploration of the putative new underlying symmetry which prevents
the dark matter particles from decaying.  The favorite symmetry at the
present time is SUperSYmmetry (SUSY), but there are other
possibilites, as summarized in the following table:

\begin{table}[h]
\begin{tabular}{cccc}
\hline
    \tablehead{1}{c}{b}{motivation}
  & \tablehead{1}{c}{b}{WIMP name}
  & \tablehead{1}{c}{b}{particle type}
  & \tablehead{1}{c}{b}{stability principle}   \\
\hline
supersymmetry & lightest SUSY particle& Majorana fermion & R-parity conservation \\
left-right asymmetries & iso-singlet neutral lepton & Majorana fermion & no mixing with neutrinos \\
extra dimensions & lightest Kaluza-Klein & fermion or boson & universal extra dimension\\
& particle&&\\
\hline
\end{tabular}
\caption{WIMP candidates motivated from particle physics}
\label{tab:b}
\end{table}

The Large Hadron Collider (LHC) will start operating in 2007; if the
underlying particle dark matter symmetry allows for new strongly
interacting particles at the TeV scale (e.g. gluinos from
supersymmetry), then the new strongly interacting particles should be
produced copiously and subsequently decay into pairs of dark matter
particles which can be detected by ``missing momentum'' measurements
in the LHC detectors.

Indirect detection of dark matter has the potential to provide crucial
information on the physics of particle dark matter which cannot be
obtained by direct detection or collider production experiments.  The
reason is that indirect detection depends upon pair annihilation of
the relic dark matter which we know must exist as an integral part of
the observed astrophysical structures in the universe, whereas direct
detection depends upon WIMP-nucleon scattering at the earth, and
collider experiments rely upon WIMP pair production.  All three
techniques are nearly completely complementary to each other in terms
of dark matter source location and dark matter interaction mechanism,
as summarized in the following table:

\begin{table}[h]
\begin{tabular}{cccc}
\hline
  & \tablehead{1}{c}{b}{WIMP location}
  & \tablehead{1}{c}{b}{WIMP interaction}
  & \tablehead{1}{c}{b}{observation methods}   \\
\hline
direct detection & Earth's surface & WIMP-nucleus scattering & underground detectors \\
colliders & irrelevant & WIMP pair production & LHC, ILC detectors\\
indirect detection & earth, sun, galaxy & WIMP pair& gamma, neutrino, anti-matter\\
&&annihilation&observatories\\
\hline
\end{tabular}
\caption{WIMP detection experiment classes}
\label{tab:c}
\end{table}

The idea of using gamma-rays to detect particle dark matter has a long
history~\cite{gammas}.  A new era in gamma-ray astronomy will be
opened with a successful launch of the Gamma-ray Large Area Space
Telescope (GLAST) in late 2007.  GLAST may be the single most
important experiment in the area of indirect detection of dark matter
for several years to come, as compared to anti-matter or neutrino
observatories, or ground based gamma-ray observatories.  The reason is
that gamma-rays have a high detection efficiency, unlike neutrinos,
and gamma-rays also travel in straight lines, unlike anti-matter,
which permits the ``imaging'' of the dark matter sources.  A large
fraction of the WIMP annihilation gamma-rays are expected to fall into
the ``sweet spot'' of the GLAST energy range $\approx1$~GeV, which is
below the threshold of ground based gamma-ray observatories.

The GLAST Large Area Telescope (LAT)~\cite{glast} consists of a
tracking detector, calorimeter, and anti-coincidence detector.  The
tracking detector is composed of 36 layers of Silicon strip detectors
interleaved with Tungsten foils, and serves as the target for
converting the gamma-rays as well as measuring the tracks of the
conversion electron-positron pair.  The calorimeter is composed of 8
layers of stacked Cesium Iodide crystal logs readout with PIN
photodiodes at both ends of each log.  The anti-coincidence detector
is composed of overlapping plastic scintillator tiles readout with
PMTs via wavelength shifting fibers.  At the time of writing of this
paper, the LAT has been accepted by NASA for integration with the
observatory.  The launch of GLAST into low earth orbit is scheduled
for fall 2007.

The LAT is designed to present an effective area of $\sim$10000 square
centimeters with a field of view covering $\sim$1/6 of the full sky.
The standard mode of operation will be a continuous scan of the sky.
The energy threshold is 20 MeV and gamma-ray energies up to 300 GeV
can be measured.  The angular resolution for the LAT is significantly
better than that of EGRET (factor of at least a few), as is the energy
resolution, especially above 10 GeV.  The mission lifetime is designed
for 5 years.  The resulting point source sensitivity ($\sim$$3\times
10^{-9}cm^{-2}s^{-1}$) is better than that of EGRET by a factor of 30
(below 10 GeV) to 100 (above 10 GeV).

If some of the gamma-rays observed by EGRET originated in pair
annihilation of WIMPs, then GLAST has an excellent chance to
independently demonstrate the existence of particle dark matter.  In
any case, because indirect detection of dark matter offers a
complementary method for probing the underlying symmetry preventing
the decay of dark matter particles, sources consistent with WIMP
annihilation radiation located by GLAST will be compelling targets for
future gamma-ray observatories.

The dark matter sources available to GLAST may be categorized into 4
classes: galactic center, galactic halo, galactic satellites, and
extragalactic.  Two gamma-ray sources found by EGRET may have
contributions from WIMP pair annihilation: 1) the ``GeV-excess'' found
in the galactic diffuse~\cite{egretgaldif}, and 2) the source at the
galactic center (3EG J1746-2851).  Of course, WIMP pair annihilation
is only one of several possible explanations, and GLAST should be able
to rule out or confirm the various hypotheses.  The hypotheses for the
galactic diffuse ``GeV-excess'' include:

\begin{itemize}
\item contribution from diffractive proton scattering~\cite{diffractive}
\item inverse Compton scattering from a larger than local cosmic ray electron flux~\cite{smr04}
\item WIMP annihilation
\end{itemize}

\begin{figure}
\includegraphics[height=.3\textheight]{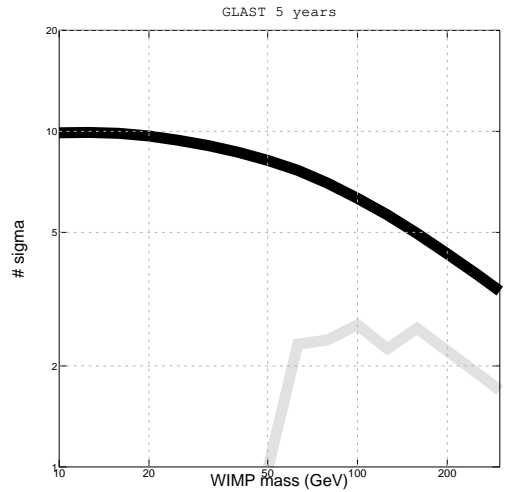}
\caption{
\#sigma vs WIMP mass for a line branching fraction of 0.1\%; grey is
for $\gamma-Z^0$ and black is for $\gamma-\gamma$.  The assumptions
are given in the text.
\label{nsigma_vs_mass-fig}}
\end{figure}

The most convincing signature for WIMP annihilation would be
$\gamma-\gamma$ and/or $\gamma-Z^0$ spectral lines.  As a
back-of-the-envelope estimate, consider the high latitude region
$|b|>$10deg ($|b|>$30deg for $|l|<$30deg), 5 years of GLAST data, and
the galactic diffuse background predicted by GALPROP within
$\Delta$E/E=0.235 at $E=M_{WIMP}$.  If we assume that tree-level WIMP
annihilation contributes to 30\% of the galactic diffuse flux for
$E>.01M_{WIMP}$ and a line branching fraction of 0.1\%, then the
significance (\# line counts/sqrt(\# bkgd counts)) as a function of
WIMP mass is shown in figure~\ref{nsigma_vs_mass-fig}.

The EGRET gamma-ray source at the galactic center~\cite{egretgc} may
have contributions from particle dark matter radiation.  The source
may or may not be coincident with the supermassive black hole; at high
energies ($>$~5GeV) the source appears to be offset~\cite{hoopding04}.
The hypotheses for the galactic center source include:

\begin{itemize}
\item cluster of pulsars
\item inverse Compton scattering from an electron source
\item accretion onto the supermassive black hole
\item WIMP annihilation
\end{itemize}

The main signatures for the WIMP annihilation hypothesis will be the
spectrum, spatial location and extent, lack of variability, and lack
of counterparts.  The most convincing signature for WIMP annihilation
will come from the energy spectrum, e.g. GLAST should be able to
search for a cutoff of the source spectrum at the mass of the WIMP.
However, due to the complex astrophysics present in the galactic
center region, we may ultimately need to observe WIMP annihilation
lines in order to lay the question to rest; this may require going
``beyond-GLAST'' due to suppression of the line branching fractions.
The results of the LHC, the next round of direct detection
experiments, and GLAST will basically shape the requirements for the
next generation of dark matter detection experiments.  Around 2010 we
should be ready to formulate the requirements for the new experiments,
whether they be on the ground, in space, and/or underground.





\bibliographystyle{aipproc}   


\end{document}